\documentclass[aps,prb,reprint,superscriptaddress]{revtex4-1}

\usepackage{graphicx}
\usepackage{dcolumn}
\usepackage{bm}
\usepackage{color}

\begin{document}


\title[Sample title]{Efficient conversion of light to charge and spin in Hall-bar microdevices}

\author{L. N\'{a}dvorn\'{i}k}
 \email{nadvl@fzu.cz}
 \thanks{These authors contributed equally to this work.}
 \affiliation{Institute of Physics ASCR, v.v.i., Cukrovarnick\'{a} 10, 16253 Praha 6, Czech Republic}
 \affiliation{Faculty of Mathematics and Physics, Charles University, Ke Karlovu 3, 12116 Praha 2, Czech Republic}
\author{J. A. Haigh}%
 \thanks{These authors contributed equally to this work.}
 \affiliation{Hitachi Cambridge Laboratory, J. J. Thomson Avenue, CB3 0HE Cambridge, UK}
\author{K. Olejn\'{i}k}
 \affiliation{Institute of Physics ASCR, v.v.i., Cukrovarnick\'{a} 10, 16253 Praha 6, Czech Republic}
\author{A. C. Irvine}
 \affiliation{Microelectronics Research Centre, Cavendish Laboratory, University of Cambridge, CB3 0HE, UK}
\author{V. Nov\'{a}k}
 \affiliation{Institute of Physics ASCR, v.v.i., Cukrovarnick\'{a} 10, 16253 Praha 6, Czech Republic}
\author{T. Jungwirth}
 \affiliation{Institute of Physics ASCR, v.v.i., Cukrovarnick\'{a} 10, 16253 Praha 6, Czech Republic}
 \affiliation{School of Physics and Astronomy, University of Nottingham, Nottingham NG7 2RD, UK}
 \author{J. Wunderlich}
 \affiliation{Institute of Physics ASCR, v.v.i., Cukrovarnick\'{a} 10, 16253 Praha 6, Czech Republic}
 \affiliation{Hitachi Cambridge Laboratory, J. J. Thomson Avenue, CB3 0HE Cambridge, UK}
 


\date{\today}

\begin{abstract}
{We report an experimental study of the direct conversion of light into electrical signals in GaAs/AlGaAs Hall-bar microdevices. Our approach based on different modulation frequencies of the intensity and polarization of the laser beam allows us to disentangle the charge and spin dependent parts of the induced electrical signal, and to link them to the incident light intensity and polarization, respectively.  We demonstrate that the efficiency of the light to spin conversion in our electrical polarimeter is strongly enhanced by adding a drift component to the transport of the spin-polarized photocarriers, as compared to a purely diffusive transport regime of the device. For a micron-size focused laser beam, the experiments demonstrate that the light to charge and spin conversion efficiency depends on the precise position of the light spot, reflecting the spatially dependent response function of the Hall cross.}



\end{abstract}

\maketitle



Within only one decade since first experimental observations\cite{kato2004,Wunderlich2004,Wunderlich2005}, the spin Hall effect   with its reciprocal counterpart, the inverse spin Hall effect\cite{Valenzuela2006,Saitoh2006,zhao2006}, have developed from subtle academic phenomena to practical tools for exploring a variety of fields in the fundamental and applied spintronics research \cite{jungwirth2012}. Spin Hall effects are now commonly used to generate and detect spin currents in non-magnetic semiconductors and metals\cite{kato2004,Wunderlich2004,Wunderlich2005,Valenzuela2006,Saitoh2006,jungwirth2012}, the spin Hall effect in heavy transition metals can trigger magnetization reversal in an adjacent ferromagnet\cite{Miron2011b,liu2012}, and the spin Hall effect was used to demonstrate a spin-transistor concept\cite{wunderlich2010} or served in electrical sub-nanosecond time resolved experiments as a THz wave radiator \cite{kampfrath2013}. It has been also proposed to apply the inverse spin Hall effect in electrical polarimeters\cite{wunderlich2009,wunderlich2010}. In these devices the degree of circular polarization of incident light is directly converted into a transverse voltage via spin-orbit interaction acting on optically generated spin-polarized photocarriers. The concept is scalable, does not require any mechanical component, and can work at room temperature.
\begin{figure}
\includegraphics[width=1.0\columnwidth]{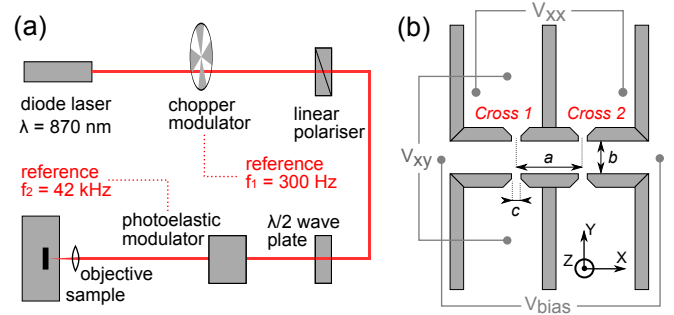}
\caption{\label{fig:sketches} (Color online) (a) The experimental setup with its two modulators: the chopper modulator of the light intensity at reference frequency $f_1=300$~Hz and the photo-elastic modulator that switches the $\sigma_+/\sigma_-$ circular polarization at $f_2=42$~kHz. (b) Sketch of the device design with Hall crosses on samples A and B. Indicated dimensions are described in the text. Inset: The spatial coordinate system.}
\end{figure}

In this letter, we study the electrical response via spin dependent Hall effects to local optical spin injection of two devices with similar Hall cross (HC) geometries. To explore the sensitivity of the devices at different experimental conditions, we perform measurements in three regimes: (i) In the diffusive regime, the spin current is generated from a non-uniform spatial distribution of spin-polarized photocarriers along the transport channel after local absorption of circularly polarized light. (ii) In a drift-dominated regime, the sensitivity is highly amplified due to the  applied longitudinal bias. 
(iii) Finally, in measurements with the laser beam focused to a $\sim 1\mu$m-size spot comparable to the HC size, the highly spatially resolved regime reveals the response function of the HC. In our experiments we employ distinct modulation frequencies of the incident light intensity and polarization. This allows us to decouple charge and spin related signals generated simultaneously by the photo-excited carriers in the Hall bar microdevice. 

%

Our experimental setup is sketched in Fig.~\ref{fig:sketches}(a). All experiments presented here were performed at room temperature (RT). A diode laser was used to generate monochromatic light of a wavelength 870~nm. When absorbed in GaAs, it creates electron-hole pairs at the edge of the bandgap, corresponding wavelength of which is $\sim$~871.7~nm at RT. The laser light is attenuated to the desired intensity and modulated at the first reference frequency $f_1=300$~Hz by a chopper wheel. The proper circular polarization state is then created by a polarizer, a $\lambda/2$-plate, and a photo-elastic modulator (PEM) working at the frequency $f_2=42$~kHz in a $\lambda/4$-regime. This sequence of components produces a light beam whose intensity is modulated at the frequency $f_1$ and the helicity of its circular polarization switches between $\sigma_+$ and $\sigma_-$ at the frequency $f_2$. An infra-red high-quality objective with 100$\times$ magnification  focuses the beam on the sample surface, allowing us to create a light spot with a minimum diameter 1~$\mu$m\footnote{The `laser spot diameter' refers to the full width at half maximum (FWHM) of its Gaussian beam profile. This value was measured experimentally by scanning with the spot over an insulating trench and determined from the dependence of the conductivity in the conductive part on the spot position.}. The objective can be positioned by a high-precision 3D piezo-electric stage which facilitates scanning of the laser spot over the device. The real time intensity variation and spot position were monitored by a silicon detector and CCD camera on laser beams separated by two beam-splitters, respectively. Finally, the electric signals were pre-amplified and sensed by a couple of lock-in amplifiers operating at frequencies $f_1$ and $f_2$.

Experiments were performed on  samples A and B that have similar layer-structure but differ in surface nanolithography. Both samples consist of a double-layer GaAs/Al$_x$Ga$_{1-x}$As, with $x=0.3$ and layer thicknesses 90/565 and 90/587 nm, respectively, grown on semi-insulating GaAs substrate. To make the GaAs layer conductive for electrical sensing/biasing, it is remotely doped by Si located in the AlGaAs layer and separated by a 8 and 15~nm undoped AlGaAs spacer from the GaAs/AlGaAs interface, respectively. The Si donor concentrations are $n_D=6\times10^{12}$ and $9\times10^{11}$~cm$^{-2}$ for sample A and B, respectively. 


The nanolithographical pattern is depicted in Fig.~\ref{fig:sketches}(b) and the corresponding dimensions for samples A and B are $a=2$, $b=1$, $c=0.30$~$\mu$m and $a=6$, $b=2$, $c=0.45$~$\mu$m, respectively. The width of channels was chosen to allow for higher current density and for comparable dimensions of the HCs and the focused laser spot. Note that in the discussion below we refer to the positions of HCs 1 and 2 as marked in Fig.~\ref{fig:sketches}(b).

We first make a remark on the optical generation of spin currents. An absorption of circularly polarized light in the bulk GaAs creates both the local increase of carrier density and the local spin polarization (it is usually referred to as the optical orientation\cite{agranovich1984}). According to the semiclassical drift-diffusion theory of coupled charge and spin currents\cite{dyakonov2007}, one can write for the charge current $\bm{j}$
\begin{equation}
\bm{j}/e=\mu n \bm{E} + D\nabla n + \gamma\mu\bm{E}\times \bm{S} + \gamma D \nabla\times\bm{S},
\label{eq:currents}
\end{equation}
where $\bm{E}$ is the vector of the applied electric field, $n$ is the electron concentration, and $\bm{S}$ is the vector of the spin polarization density. Constants $e$, $\mu$, $D$, and $\gamma$ are the absolute value of the electron charge, the electron mobility, the diffusion coefficient, and the spin Hall angle, respectively. The first two terms in Eq.~\ref{eq:currents} represent the standard charge drift-diffusion equation. The third term corresponds to the anomalous Hall effect \cite{nagaosa2010}. The fourth term  describes the inverse spin Hall effect\cite{Valenzuela2006,Saitoh2006} if a charged diffusive current is absent, i.e. in the case  of the pure spin current in the system. We distinguish this from the situation in which a polarized charge diffusive current, e.g., generated by optical excitation, leads to a charge transverse current which we associate here with a regime closer to the anomalous Hall effect. This distinction is made more clear by the fact that the spin Hall effect has a precise definition of a pure spin current being generated by a charge current and, therefore its inverse is associated with a pure spin current generating a transverse charge current. 
\begin{figure}
\includegraphics[width=1.0\columnwidth]{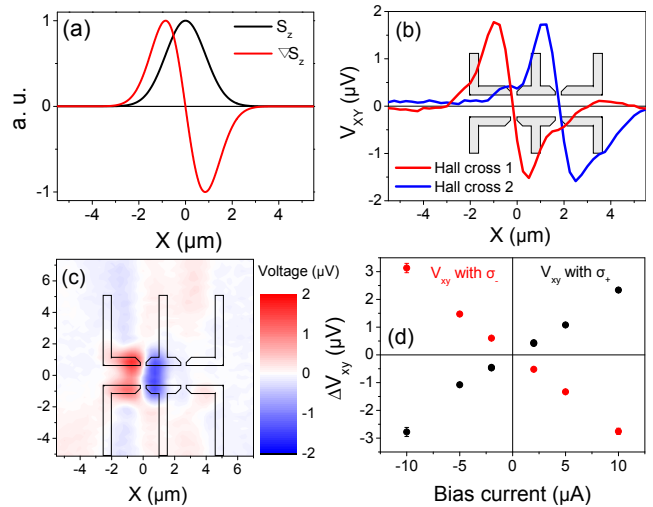}
\caption{\label{fig:diffusion} (Color online) Diffusive spin currents in sample A. (a) Simulated Gaussian profile of the optically injected spin polarization density $S_z$ with FWHM~$=2\ \mu$m (black curve) and the corresponding diffusive spin current proportional to $\nabla S_z$ (red curve). (b) Experimental Hall voltages measured at $f_2$ and HC 1 (red curve) and HC 2 (blue curve) as a function of a 1D displacement of the light spot (FWHM~$\approx2\ \mu$m and $P=15\ \mu$W). (c) The Hall voltage measured at $f_2$ and HC 1 with respect to the 2D position of the spot (same parameters as in (b)). (d) A typical dependence of the Hall voltage at HC 1 on the drift current applied along the transport channel for $\sigma_+ / \sigma_-$ helicities and for power $P=700$~nW.}
\end{figure}

Absorption of a circularly polarized light at normal incidence generates both the non-equilibrium charge concentration $n(x,y)$ and spin polarization $S_z(x,y)$, the inhomogeneous distribution of which can be approximated by a 2D Gaussian function in the $(x,y)$-plane. The non-uniform $S_z(x,y)$ leads to a non-zero diffusive spin current which drives the spin dependent Hall effect described by the last term in Eq.~\ref{eq:currents}. The drift term proportional to the vector product of the applied electric field and $S_z(x,y)$ describes the anomalous Hall contribution to the transverse charge current. 

We proceed now to experimental results on sample A in the diffusion dominated regime. Without an applied electric bias along the channel, the voltage measured between the Hall contacts (see Fig.~\ref{fig:sketches}(b)) compensates the equilibrium current $j_y = \gamma D (\nabla S_z)_x$. Fig.~\ref{fig:diffusion}(a) shows a simulated profile of the light spot of width 2~$\mu$m (black line) and the corresponding simulated Hall signal (red line), which is proportional to the spatial derivative of $\bm{S}$, assuming that a point-like HC is moved with respect to the light spot. Corresponding experimental data, shown in Fig.~\ref{fig:diffusion}(b), were measured at the reference frequency $f_2$ by scanning a 2~$\mu$m wide spot ($2\times$ the width of the transport channel) of laser power $P=15\mu$W across two HCs of  sample A at normal incidence. As the transverse
Hall voltages $V_{xy}$ are sensed at a frequency corresponding to the helicity oscillation between $\sigma_+/\sigma_-$, we detect only the signals related to the spin dependent Hall effect. The voltages measured on the left (right) HC with respect to the spot position are shown by the red (blue) curve. Note that the signals reflect the expected spatial dependences and that the positions of zero Hall signals, located in centers of the HCs, are separated by  2~$\mu$m. This is in agreement with the design of this sample, as seen Fig.~\ref{fig:sketches}(a). (Note that a similar observation was previously reported by X.~W.~He \textit{et al.}\cite{he2008} on Al$_x$Ga$_{1-x}$N/GaN heterostructures at RT on a $\sim1000\times$ larger scale.) In Fig.~\ref{fig:diffusion}(c) we show a 2D scan of the Hall voltage on the left HC. It again shows the inversion-asymmetric signal centered at the HC.

\begin{figure}
\includegraphics[width=1.0\columnwidth]{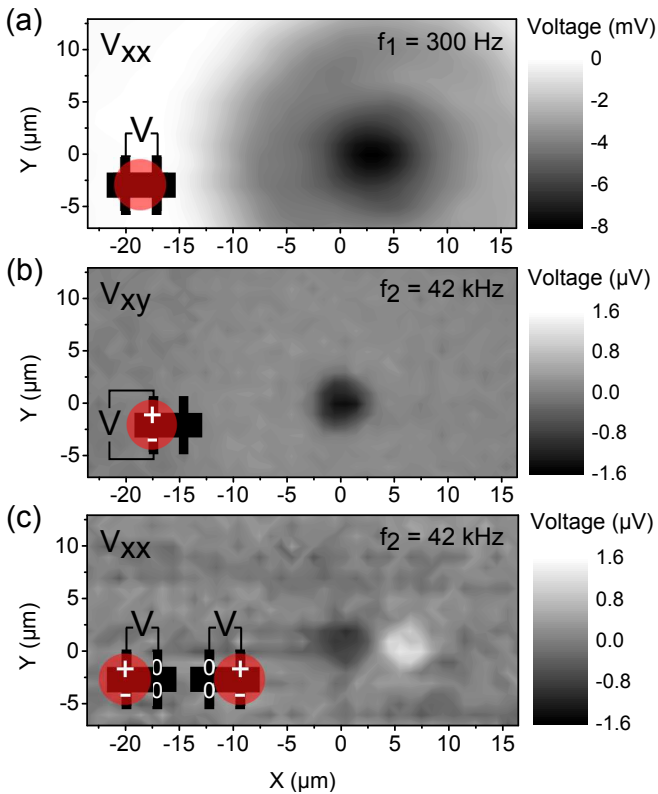}
\caption{\label{fig:polarimeter} (Color online) Drift-dominated spin currents in sample B. The sample is biased at $I=10\ \mu$A and scanned with a light spot of size 5~$\mu$m ($2.5\times$ the width of the channel) and power $P=80$~nW. (a) Variation in the longitudinal voltage $V_{xx}$  measured at chopper reference frequency $f_1$. (b) The Hall voltage $V_{xy}$ sensed by the HC 1 at PEM frequency $f_2$. (c) The longitudinal voltage $V_{xx}$ measured at PEM frequency $f_2$ between HC 1 and 2. Insets: Sketches of polarities of Hall voltages with respect to the spot position (red circles).}
\end{figure}
The drift component of the spin polarized charge current, described by the third  term in Eq.~\ref{eq:currents}, adds up to the diffusive term when applying electric bias along the channel. The relative contribution of the diffusive and drift components to the detected spin dependent Hall signal varies with the laser power and the applied electric bias. A fully drift-dominated regime can be achieved by attenuating the laser power 20$\times$ to 700 nW, and by applying a sufficiently large electric bias, as shown in Fig.~\ref{fig:diffusion}(d). A linear dependence of the Hall voltage on the bias current is measured in this plot at the point of maximal diffusive signal at $P=700$~nW. The absence of any off-set from the origin gives an experimental evidence that the diffusive contribution is completely suppressed under these conditions. The measured linear behavior up to 10 $\mu$A of bias current confirms that the experiment is in the linear-response regime described by Eq.~\ref{eq:currents} and indicates that the sample remains below the saturation threshold. Indeed, the DC electric field between the left and right HCs that corresponds to maximal bias current $I=10\ \mu$A was $\sim 150$~V/cm, compared to the much higher saturating electric field of  $E_{sat}\sim1$~kV/cm reported for low doped bulk GaAs \cite{miah2007}. 

Figs.~\ref{fig:polarimeter}(a-c) show experimental results for sample B and $I=10\ \mu$A. Here the channel is scanned by a defocused laser beam of a diameter $5\ \mu$m (which is $2.5\times$ larger than the width of the channel) and of low power $P=80$~nW. In Fig.~\ref{fig:polarimeter}(a), the longitudinal voltage drop $V_{xx}$ between HC 1 and 2 is measured at the chopper reference frequency $f_1$. The signal is therefore correlated with the photoexcited carriers that contribute to the variation of the conductivity between the HCs.
The Hall signal (the transverse voltage $V_{xy}$) detected on HC 1 at the PEM frequency $f_2$ is displayed in Fig.~\ref{fig:polarimeter}(b). The response is clearly spatially symmetric which confirms the dominant drift component of the spin current. Moreover, the signal reflects the 2D Gaussian function of the light intensity distribution without any inner structure and its size roughly corresponds to the FWHM of the light spot. Thus we can conclude that the HC is acting as a point detector for this FWHM/HC size ratio.

A control experiment is shown in Fig.~\ref{fig:polarimeter}(c). A longitudinal voltage between HC~1 and 2 was detected at $f_2$ while the spot was scanning the sample surface. In this case, the data show two extrema with opposite sign. When the spot is generating locally the Hall signal at HC~1, HC~2 is in dark and thus the potential at HC~2 is roughly zero. If the spot is placed on top of  HC~2, the measured longitudinal voltage switches sign. Also consistently, the separation of the two extrema ($\approx$ 6~$\mu$m) corresponds to the separation of HCs~1 and 2 (see Fig.\ref{fig:sketches}(b)).

Note that measurements in the drift dominated regime in  bulk semiconductor and in quantum wells have been also recently reported in Ref.~\onlinecite{miah2007,vasyukov2010}. Our results highlight  that the transition from the diffusive to drift regime can dramatically amplify the sensitivity of the polarimeter device based on the spin dependent Hall effect. In particular, the unfavorable anti-symmetric shape of the signal with vanishing Hall voltage for spot centered at the HC, which is characteristic of the diffusive regime, is absent when applying the strong drift current. 

\begin{figure}
\includegraphics[width=1.0\columnwidth]{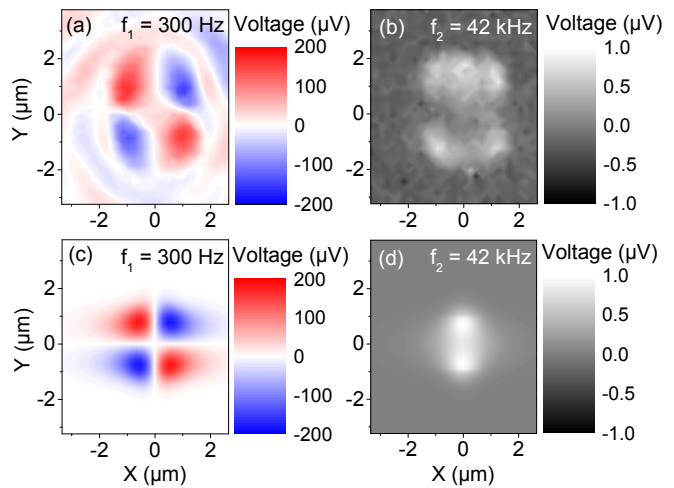}
\caption{\label{fig:respfun} (Color online) Response functions in sample B. (a) The Hall voltage $V_{xy}$ sensed at reference frequency $f_1$, related to photoinduced changes in conductivity. (b) The Hall voltage $V_{xy}$ sensed at PEM frequency $f_2$. Data in panels (a,b) are measured with the spot of diameter $1\ \mu$m and power 80~nW, the channel is biased with $I=10\ \mu$A. (c,d) Simulations of Hall voltages  at frequencies $f_1$ and $f_2$ for the same size of the HC and laser spot.}
\end{figure}

The apparently homogeneous and spherically symmetric response of the HC at defocused light spot, discussed in the previous paragraphs, changes and reveals its inner structure when the spot size is decreased by focusing it to $1/2$ of the width of the channel. In this spatially sensitive regime we resolve regions with higher Hall response which correlate with the simulated response function of the HC. 
In the experiment, the channel of sample B was biased with the current $I=10\ \mu$A and one of the HCs was scanned by the spot of a diameter $1\ \mu$m and laser power $P=80$~nW, with the spatial step size 200~nm. The transverse voltage $V_{xy}$ measured at this HC at chopper and PEM reference frequencies $f_1$ and $f_2$ is shown in Fig.~\ref{fig:respfun}(a,b). The signal at frequency $f_1$ reflects local photo-induced changes in the carrier density and thus changes in the photo-conductivity. This affects locally the current density and creates asymmetric potential distribution across the channel and, therefore, produces a non-zero transverse voltage $V_{xy}$. The effect is strongly dependent on the position where the conductivity is changed with respect to the HC design and increases at its corners. 

Note that a similar observation was made in doped quantum wells using conventional non-magnetic STM tips as local electric gates\cite{nabaei2013,folks2009}. Our experimental finding is also in agreement with a numerical simulation, the output of which is shown in Fig~\ref{fig:respfun}(c). A Poisson solver is used here to calculate the potential difference at Hall contacts using device parameters of our experiment. Each pixel in the simulation represents the result of the Poisson solver when the conductivity is varied around this point by a Gaussian function which causes an overall change of the conductivity by 10~\%, as observed in experiment. The FWHM of the Gaussian function is set to 1~$\mu$m. We observe that not only the two-fold symmetry but also the approximate size of the features agree with the experiment. The outer rings in the experimental data, not seen in theory,  are likely due to light scattered at trenches or due to complex interference effects.

The signal measured at frequency $f_2$ (see Fig.~\ref{fig:respfun}(b)) is correlated only with the spin-dependent Hall signal. It reveals the inhomogeneous response of the HC, confirming that the laser spot of size of 1/2 of the channel width gives us a sufficient resolution to measure optically the HC response function. The data also highlight that in a polarimeter device with highly focused incident beams one would need to take the HC response function into account in order the maximize its sensitivity. 

Apart from higher order complexities, the measured two-fold signal shown in Fig.~\ref{fig:respfun}(b) is in agreement with our simulations (see Fig.~\ref{fig:respfun}(d)). Again the same scanning Gaussian profile was used in the modelling. However, in order to simulate the spin dependent Hall effect, we included the  spin polarization density $S_z$ with the same Gaussian distribution as considered for the charge distribution. From the measured data and from the expression relating the charge current ${\bf j}$ and spin current ${\bf j^{s,z}}$ in the drift regime,  $j_y=\gamma j^{s,z}_x=e\gamma j_xS_z$, we infer the anomalous (spin) Hall angle $\gamma\approx2\times10^{-3}$. This value agrees with the previously reported  angles for bulk GaAs\cite{matsuzaka2009,garlid2010,ehlert2012}. 

Our experiments revealing the HC response function can be viewed as an optical spintronic  analogue of the previously employed magnetic force microscopy (MFM) technique\cite{nabaei2013,folks2009,baumgartner2006,thiaville1997}. In the latter approach, a local flux of magnetic field and the ordinary Hall effect were used instead of our local optical spin injection and the relativistic, spin dependent Hall effects. We point out that our optical experiments are broadly consistent with these earlier MFM measurements. We also recall, however, that in our case the resolution  is limited by the wavelength of the incident light. On the other hand, in the MFM experiments, the magnetic tips are usually electrostatically coupled to the conductive channel which adds additional signals due to the affected current distribution. This problem is circumvented in our optical measurements by using the distinct modulation frequencies for the light intensity and polarization, and the lock-in detection of the respective charge and spin dependent signals. 

In conclusion, we have demonstrated that transverse electrical signals generated by polarized light via spin dependent Hall effects can be enhanced by more than two orders of magnitude with an applied drift current along the transport channel of the Hall bar microdevice.  In experiments with a focused laser beam we have complemented this observation with the detection of the spatially dependent sensitivity of our polarimeter which reflects the HC response function. Finally, our double-frequency modulation technique allows to convert information coded simultaneously in the intensity and polarization of the incident light into separate charge and spin dependent electrical signals. All these observations may find utility in designing spin Hall based convertors, modulators,  polarimeters or other related opto-spintronic microdevices.

\begin{acknowledgments}
We acknowledge  support from the Charles University grant SVV-2014-260094, from the European Research Council (ERC) advanced grant no. 268066, from the Ministry of Education of the Czech Republic grant no. LM2011026, from the Grant Agency of the Czech Republic grant no. 14-37427G, from the Academy of
Sciences of the Czech Republic Praemium Academiae, and from EMRP IND08-REG1 which is jointly funded
by the EMRP participating countries within EURAMET and the E.U. 
\end{acknowledgments}
 

%
%

\bibliography{bibliography1,bibliography2}

\end{document}